\documentclass[final,5p,times,twocolumn]{elsarticle}

\usepackage{lineno}
\usepackage[hidelinks]{hyperref}

\usepackage{graphicx}
\usepackage{amsmath} 
\usepackage[separate-uncertainty = true]{siunitx}
\usepackage[dvipsnames]{xcolor}
\def\pmbanner{{\hrule height 1 pt}\vskip35pt{NIMA POST-PROCESS BANNER TO BE REMOVED AFTER FINAL ACCEPTANCE}\vskip35pt{\hrule height 4pt}\vskip20pt}

\begin{document}

\begin{frontmatter}

\title{\pmbanner TCAD modeling of radiation-induced defects in 4H-SiC diodes}
\author[1,2]{Philipp Gaggl}
\author[1]{Jürgen Burin}
\author[1,2]{Andreas Gsponer}
\author[1]{Simon Emanuel Waid}
\author[1]{Richard Thalmeier}
\author[1]{Thomas Bergauer\corref{cor1}}\ead{thomas.bergauer@oeaw.ac.at}

\cortext[cor1]{Corresponding author}

\affiliation[1]{organization={Institute of High Energy Physics of the Austrian Academy of Sciences}, 
                 addressline={Nikolsdorfer Gasse 18},
                 postcode={1050}, 
                 city={Vienna}, 
                 country={Austria}}
\affiliation[2]{organization={TU Wien}, 
                 addressline={Wiedner Hauptstr. 8–10},
                 postcode={1040}, 
                 city={Vienna}, 
                 country={Austria}}

\begin{abstract}
4H silicon carbide (SiC) has several advantageous properties compared to silicon (Si) making it an appealing detector material, such as a larger charge carrier saturation velocity, bandgap, and thermal conductivity. While the current understanding of material and model parameters suffices to simulate unirradiated 4H-SiC using TCAD software, configurations accurately predicting performance degradation after high levels of irradiation due to induced traps and recombination centers do not exist. Despite increasing efforts to characterize the introduction and nature of such defects, published results are often contradictory. This work presents a bulk radiation damage model for TCAD simulation based on existing literature and optimized on measurement results of neutron-irradiated 4H-SiC pad diodes. Experimentally observed effects, such as flattening of the detector capacitance, loss of rectification properties, and degradation in charge collection efficiency, are reproduced. The EH$_4$ center is suggested as a major lifetime killer in 4H-SiC, while the still controversial assumption of the EH$_\text{6,7}$ deep-level being of donor type is reinforced.
\end{abstract}

\begin{keyword}
Silicon carbide \sep Radiation damage to detector materials (solid state) \sep TCAD \sep Numerical simulations \sep Defect states
\end{keyword}

\end{frontmatter}


\section{Introduction}\label{sec:intro}
Future high-energy physics (HEP) experiments ought to sustain unprecedented radiation fluences up to $10^{18}\:\text{n}_{\text{eq}}/\si{\centi\metre\squared}$ ($1\:\si{\mega\electronvolt}$ neutron equivalent according to NIEL hypothesis \cite{NIEL}) at innermost detector layers, while low material budget requirements forbid materials like cooling pipes. Due to its unique properties and recent improvements in manufacturing, research in 4H silicon carbide (4H-SiC) as high-energy particle detector material has been rekindled  \cite{neutron_detection, review_sic}. A venture necessitating simulations via technology computer-aided design (TCAD) software. Despite a multitude of studies on performance degradation of, and defect formation in 4H-SiC after irradiation, no TCAD model to accurately reproduce these findings exists to the best of our knowledge. Utilizing the TCAD software Sentaurus \cite{URL_Sentaurus}, this work represents our first results towards developing a comprehensive radiation damage model for 4H-SiC up to high irradiation fluences. Focused solely on bulk defects, literature defect parameters are adapted to reproduce measurements on neutron-irradiated 4H-SiC samples described in \cite{Gsponer_2023}.

\section{Materials and methods}\label{sec:methods}
Five planar, $3\times 3\:\si{\milli\metre\squared}$ 4H-SiC PiN-diodes with a $50\:\si{\micro\metre}$, high resistive ($20\:\si{\ohm\centi\metre}$), epitaxial layer as active detecting region, manufactured and developed at IMB-CNM-CSIC \cite{Rafi2018_samples}, were studied. Four of them were neutron-irradiated at the TRIGA Mark II reactor at the Atominstitut in Vienna at different fluences ($\Phi$), ranging from $5\times 10^{14}\:\text{n}_{\text{eq}}/\si{\centi\metre\squared}$ to $1\times 10^{16}\:\text{n}_{\text{eq}}/\si{\centi\metre\squared}$. Detailed information can be found in \cite{Gaggl2022}. The electrical characterization included current (I-V) and capacitance (C-V), as well as charge collection efficiency (CCE) measurements of $\alpha$-particles in forward and reverse operating bias \cite{Gsponer_2023}.\\
Due to the p\textsuperscript{++}-implant and metalization covering the active diode area, the proposed TCAD model considers only bulk defects, neglecting contributions of interface defects and oxide charges. Results of existing literature characterizing radiation-induced defects in 4H-SiC are subject to differing manufacturers and material quality, thus widely dispersed and sometimes contradictory. The ensuing large parameter space served as a starting point for the optimization of individual values to best fit experimental results from \cite{Gsponer_2023}. The investigated publications focus on neutron and electron-irradiated 4H-SiC since results compare very well \cite{Hazdra2014, Hazdra2019, Rafi2020}. Active defects in the simulation are either intrinsic (D \& B) or radiation-induced (Z$_{1,2}$, EH$_{6,7}$ \& EH$_4$) via a linear introduction rate ($N_i=f_i\cdot\Phi$). Defect types and cross-sections are assumed constant over the whole range of neutron fluences. Due to the low dark current levels of 4H-SiC, reverse I-V measurements are dominated by electric noise and surface currents, restricting TCAD comparison to forward bias operation. I-V simulations have been conducted in \textit{SDevice}, using an AC Analysis in \textit{Single-Device Mode} to simultaneously obtain C-V data ($10\:\si{\kilo\hertz}$ AC-frequency \cite{Gsponer_2023}) at equal simulation settings. Obtained field constellations were saved in $50\:\si{\volt}$ steps and used to simulate detection performance, utilizing the \textit{HeavyIon} model to simulate the depth-dependent energy loss of $\alpha$-particles corresponding to the source used in \cite{Gsponer_2023}. Applied settings and physics models are described in \cite{URL_RD50_talk, URL_Synopsys_training1, URL_Synopsys_training2}.

\section{Results}\label{sec:results}
\autoref{tab:results} summarizes the optimized defects of the TCAD irradiation model and their respective parameters, as well as publications best fitting the final obtained values. Aside from the widely known \textit{major lifetime killers} in 4H-SiC (Z$_{1,2}$ \& EH$_{6,7}$), the EH$_4$ defect cluster proved to be crucial as well. As discussions about the type of the EH$_{6,7}$ defect continue, the presented model strongly suggests it to be of donor-type \cite{Booker2016, Bathen2024}. Aside from slight improvements in C-V conformity, the intrinsic Boron centers (B \& D) do not affect results.

\begin{table}[ht]
\vspace{-0.4cm}
\centering
\caption{Optimized TCAD model parameters: Defect type, activation energy, e$^-$/h$^+$ capture cross-sections, and concentration after $1\:\si{\mega\electronvolt}$ equivalent neutron fluence $\Phi$. $E_C$ and $E_V$ denote the conduction and valence band energy. Additionally, the investigated literature that best fits the respective value, is given.}
\vspace{0.01\textwidth}
\label{tab:results}
\resizebox{0.48\textwidth}{!}{
\begin{tabular}{r|l|l|l|l|l}
Defect & Type & $E\:[\si{\electronvolt}]$ & $\sigma_e\:[\si{\centi\metre\squared}]$ & $\sigma_h\:[\si{\centi\metre\squared}]$ & $N\:[\si{\per\centi\metre\cubed}]$
\\ \hline
Z$_{1,2}$ & Acceptor & $E_C-0.67$ \cite{Klein2009} & $2.0\cdot 10^{-14}$ \cite{Klein2009} & $3.5\cdot 10^{-14}$ \cite{Klein2006} & $5.0\cdot\Phi$ \cite{Hazdra2014} \\ \hline
EH$_{6,7}$ & Donor & $E_C-1.60$ \cite{Knezevic2023} & $9.0\cdot 10^{-12}$ \cite{Knezevic2023} & $3.8\cdot 10^{-14}$ \cite{Knezevic2023} & $1.6\cdot\Phi$ \cite{Hazdra2014} \\ \hline
EH$_4$ & Acceptor & $E_C-1.03$ \cite{Alfieri2005} & $5.0\cdot 10^{-13}$ \cite{Booker2014} & $5.0\cdot 10^{-14}$ \cite{Hazdra2014} & $2.4\cdot\Phi$ \cite{Hazdra2014} \\ \hline
B & Donor & $E_V+0.28$ \cite{Capan2022} & $2.0\cdot 10^{-15}$ \cite{Klein2011} & $2.0\cdot 10^{-14}$ \cite{Hazdra2014} & $1.0\cdot 10^{14}$ \\ \hline
D & Donor & $E_V+0.54$ \cite{Capan2022} & $2.0\cdot 10^{-15}$ \cite{Klein2011} & $2.0\cdot 10^{-14}$ \cite{Hazdra2014} & $1.0\cdot 10^{14}$ 
\end{tabular}}
\end{table}

\begin{figure}[ht]
\centering
\includegraphics[width=0.46\textwidth]{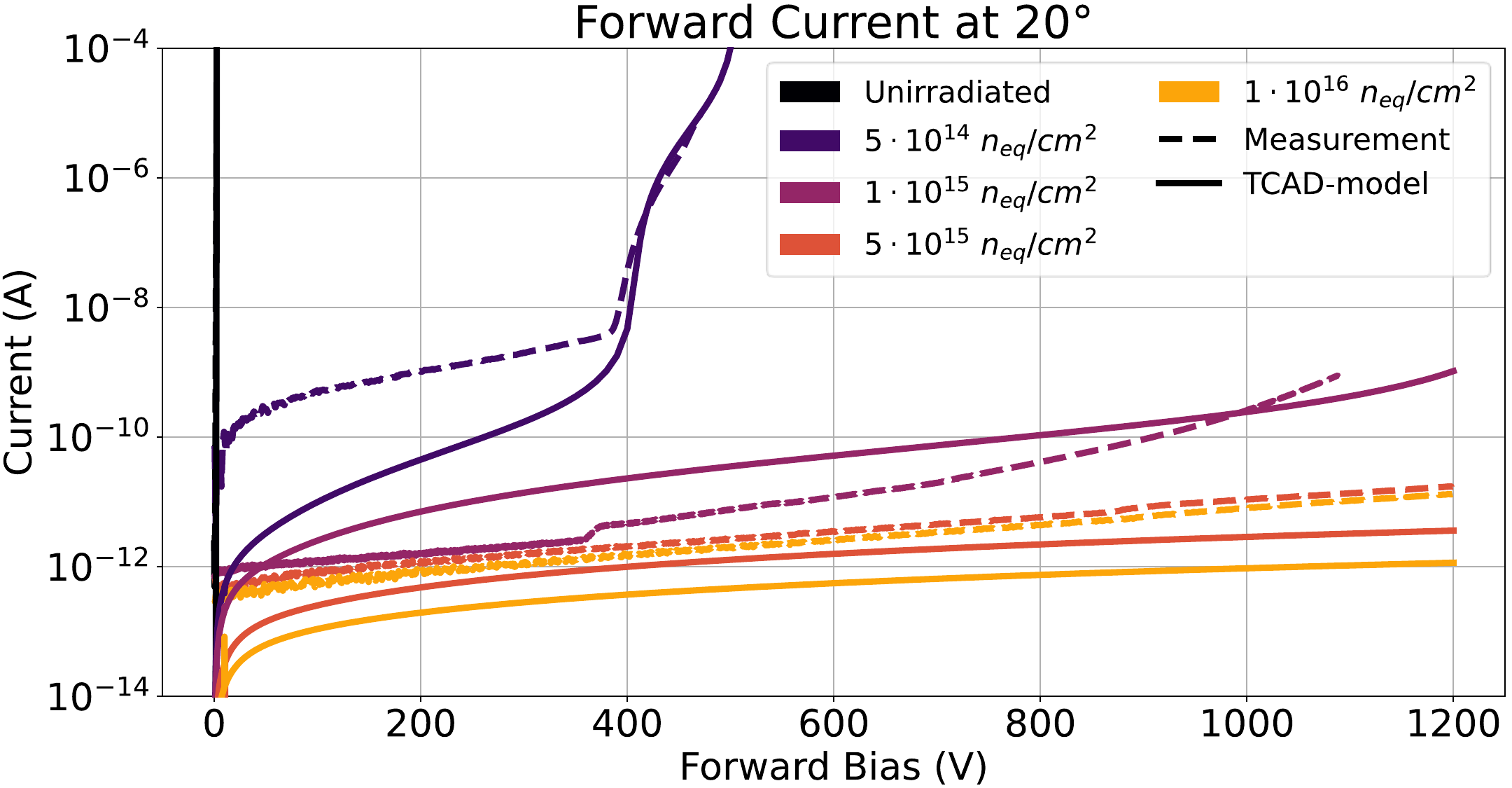}
\vspace{-0.02\textwidth}
\caption{Forward bias I-V measurements \cite{Gsponer_2023} vs. simulation.}
\label{fig:IV}
\end{figure}

\begin{figure}[ht]
\centering
\includegraphics[width=0.46\textwidth]{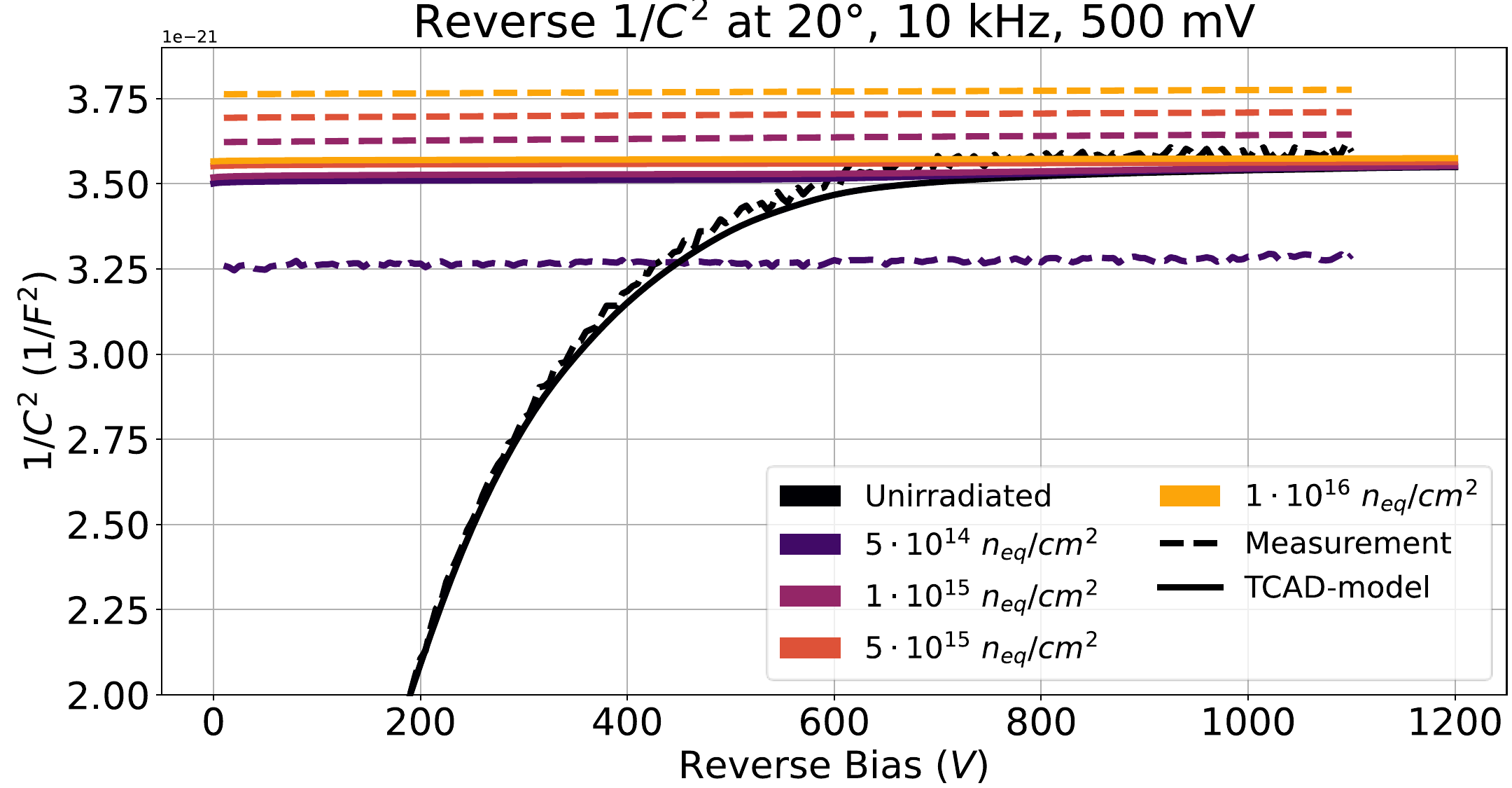}
\vspace{-0.02\textwidth}
\caption{Reverse bias 1/C$^2$-V measurements \cite{Gsponer_2023} vs. simulation. The unirradiated case (black) represents the characteristic depletion of a PiN-diode under reverse operation, the extracted full depletion voltage is $325\:\si{\volt}$.}
\label{fig:CV}
\end{figure}

\begin{figure}[ht]
\centering
\includegraphics[width=0.46\textwidth]{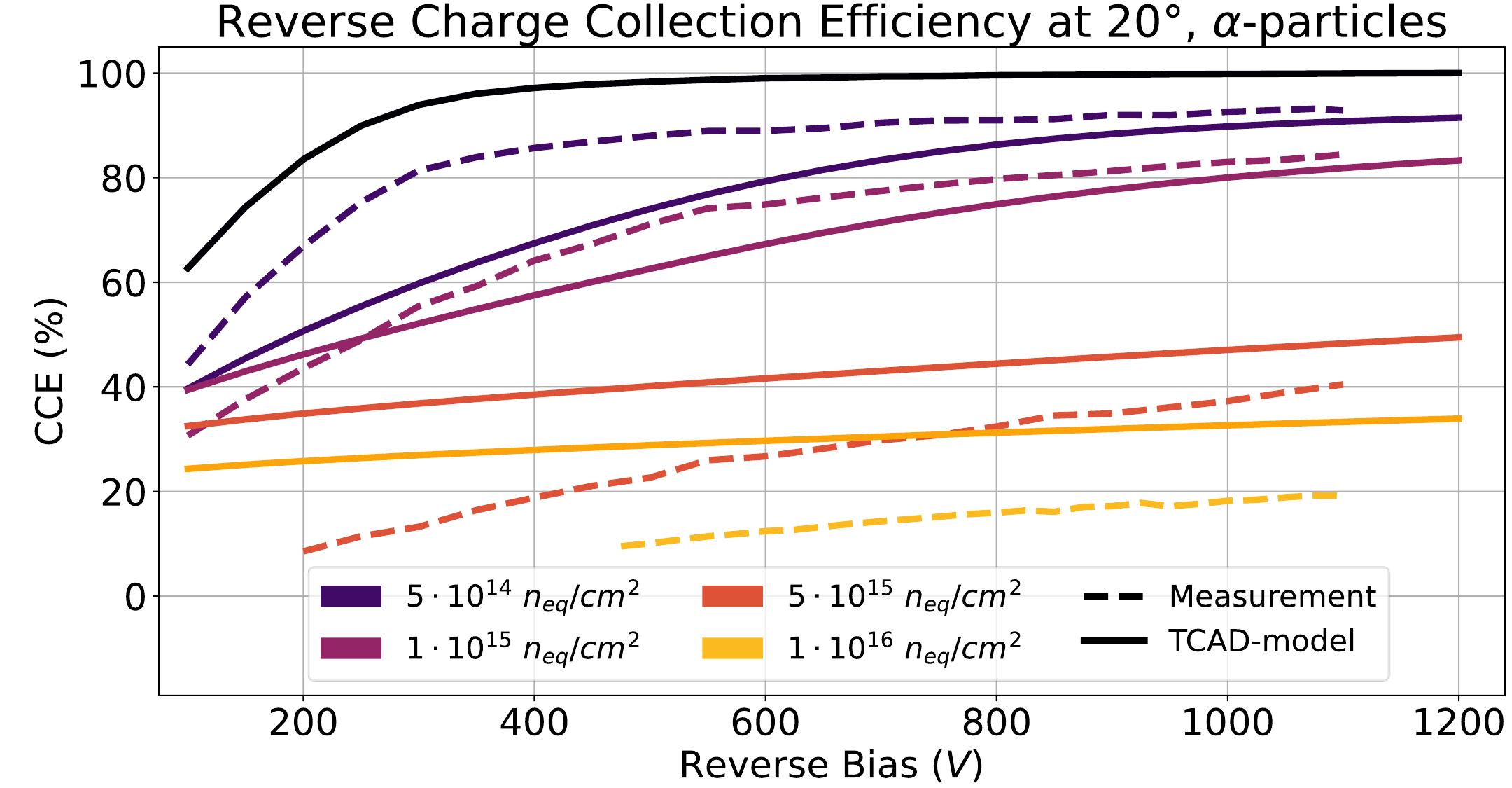}
\vspace{-0.02\textwidth}
\caption{Reverse bias CCE measurements \cite{Gsponer_2023} vs. simulation.}
\label{fig:CCE}
\end{figure}

\autoref{fig:IV} depicts measured and simulated I-V results in forward bias. The model accurately predicts the 4H-SiC device losing its rectifying properties with increasing irradiation, allowing for forward biasing up to high voltages \cite{Gsponer_2023, Rafi2020}. Absolute values are in good agreement with experiments, despite increasingly underestimating current levels with rising fluence, which may be attributed to surface currents omitted in the model. Simulations suggest a high concentration of trapped holes (EH$_{6,7}$) near the top, and electrons (EH$_4$) near the bottom due to their respective abundance within the p$^{++}$ implant and the n$^+$ buffer layer \cite{Rafi2018_samples}. The resulting potential forms an electric field barrier that suppresses current flow and increases with the irradiation fluence. The current level in this confinement is mainly determined by the majority carrier trap (Z$_{1,2}$ \& EH$_4$) concentration. At sufficient bias, defects on either side become fully occupied, increasing the respective carrier lifetimes and restoring conducting properties, as can be observed for the lowest fluence in \autoref{fig:IV}. The resulting rectification voltage is determined by the EH$_4$ concentration, as it reaches full occupations earlier. Not displayed are reverse I-V simulations, which show a slight increase in leakage current with irradiation fluence that remains below an order of magnitude increase. Although in agreement with other studies \cite{Gaggl2022, Rafi2020}, no further conclusion can be made, as simulated currents are multiple orders of magnitude below the experimental measurement limitations of around $100\:\si{\femto\ampere}$.
\\The detector capacitance has been shown to \textit{flatline} after irradiation, adopting a constant value across forward and reverse bias \cite{Gsponer_2023, Rafi2020}. \autoref{fig:CV} displays measured and simulated reverse 1/C$^2$ data. While defect parameters were adjusted mainly to agree with I-V measurements, C-V simulations reproduce experimental results, with slight deviations most likely originating inaccuracies in the simulated initial doping profile.\\
\\Measured and simulated charge collection efficiency (CCE), obtained by scaling signal areas to that of the unirradiated counterpart, are shown in \autoref{fig:CCE}. The radiation-induced performance degradation can be reproduced well at low irradiation, while deviations increase towards higher fluences. 

\section{Conclusions and Outlook}\label{sec:conclusion}
A first step towards a comprehensive TCAD model for radiation-induced bulk defects in 4H-SiC has been developed. Simulated data agrees well with measurements on neutron-irradiated 4H-SiC diodes. While absolute I-V values deviate with increasing irradiation fluence, the loss of rectification properties after irradiation can be reproduced. Suggested to be originating from major shifts within the internal electric field structure due to charge separation, the deep-level defects EH$_4$ and EH$_{6,7}$ are identified as driving forces behind this process. Furthermore, the model reinforces recent and still discussed proposals of the EH$_{6,7}$ center being of donor type while also introducing the EH$_4$ defect as a significant lifetime killer in 4H-SiC, next to the established Z$_{1,2}$ center. Other trends, such as a flattening detector capacitance and a degradation in charge collection efficiency after irradiation, are recreated and agree well with experimental results at lower fluences, increasingly deviating at higher fluences. Further and more extensive irradiation studies on various types of 4H-SiC devices are needed to confirm and improve the presented simulation model, as well as to expand it towards interface defects and oxide charges. Additional 4H-SiC samples, currently under production, are scheduled for an extensive irradiation campaign, which will include various irradiation sources and is planned to cover a wide fluence range of $10^{13}\:\text{n}_{\text{eq}}/\si{\centi\metre\squared}$ to $10^{18}\:\text{n}_{\text{eq}}/\si{\centi\metre\squared}$.

\section*{Acknowledgements}\label{sec:funding}
This project has received funding from the Austrian Research Promotion Agency FFG, grant numbers 883652 and 895291. Production and development of the 4H-SiC samples was supported by the Spanish State Research Agency (AEI) and the European Regional Development Fund (ERDF), ref. RTC-2017-6369-3.




\bibliographystyle{elsarticle-num-names} 
\bibliography{bibfileTemplate}

\end{document}